\documentclass[fleqn,usenatbib]{mnras}

\usepackage{newtxtext,newtxmath}
\usepackage[T1]{fontenc}

\DeclareRobustCommand{\VAN}[3]{#2}
\let\VANthebibliography\thebibliography
\def\thebibliography{\DeclareRobustCommand{\VAN}[3]{##3}\VANthebibliography}

\usepackage{graphicx}	% Including figure files
\usepackage{amsmath}	% Advanced maths commands
\usepackage{caption}
\usepackage{subcaption}

\def\ORVOEBVHMRVM{$3.04(0.04)$}

\def\ORVOEBVLMRVM{$3.04(0.04)$}

\def\ORVOEBVHMEBV{$0.079(0.001)$}

\def\ORVOEBVLMEBV{$0.079(0.001)$}

\def\ORVOEBVHMRVS{$0.74(0.02)$}

\def\ORVOEBVHMRVMN{46(92\%)}
\def\ORVOEBVLMRVMN{46(92\%)}
\def\ORVOEBVHMEBVN{50(100\%)}
\def\ORVOEBVLMEBVN{50(100\%)}
\def\ORVOEBVHMRVSN{36(72\%)}

\def\TRVOEBVHMRVM{$3.08(0.04)$}

\def\TRVOEBVLMRVM{$2.11(0.03)$}

\def\TRVOEBVHMEBV{$0.079(0.001)$}

\def\TRVOEBVLMEBV{$0.076(0.001)$}

\def\TRVOEBVHMRVS{$0.75(0.03)$}

\def\TRVOEBVLMRVS{$0.50(0.03)$}

\def\TRVOEBVHMRVMN{47(94\%)}
\def\TRVOEBVLMRVMN{47(94\%)}
\def\TRVOEBVHMEBVN{48(96\%)}
\def\TRVOEBVLMEBVN{47(94\%)}
\def\TRVOEBVHMRVSN{36(72\%)}
\def\TRVOEBVLMRVSN{48(96\%)}
\def\TRVTEBVMBHMRVM{$3.00(0.02)$}

\def\TRVTEBVMBLMRVM{$2.12(0.02)$}

\def\TRVTEBVMBHMEBV{$0.126(0.001)$}

\def\TRVTEBVMBLMEBV{$0.081(0.001)$}

\def\TRVTEBVMBHMRVS{$0.63(0.02)$}

\def\TRVTEBVMBLMRVS{$0.41(0.02)$}

\def\TRVTEBVMBHMRVMN{47(94\%)}
\def\TRVTEBVMBLMRVMN{47(94\%)}
\def\TRVTEBVMBHMEBVN{49(98\%)}
\def\TRVTEBVMBLMEBVN{47(94\%)}
\def\TRVTEBVMBHMRVSN{36(72\%)}
\def\TRVTEBVMBLMRVSN{46(92\%)}
\def\TRVTEBVHMRVM{$2.90(0.04)$}

\def\TRVTEBVLMRVM{$1.96(0.04)$}

\def\TRVTEBVHMEBV{$0.125(0.002)$}

\def\TRVTEBVLMEBV{$0.079(0.001)$}

\def\TRVTEBVHMRVS{$0.68(0.03)$}

\def\TRVTEBVLMRVS{$0.42(0.04)$}

\def\TRVTEBVHMRVMN{46(92\%)}
\def\TRVTEBVLMRVMN{46(92\%)}
\def\TRVTEBVHMEBVN{45(90\%)}
\def\TRVTEBVLMEBVN{49(98\%)}
\def\TRVTEBVHMRVSN{36(72\%)}
\def\TRVTEBVLMRVSN{48(96\%)}
\def\CCNOCUTHMRVM{$3.21(0.07)$}

\def\CCNOCUTLMRVM{$3.94(0.06)$}

\def\CCNOCUTHMEBV{$0.166(0.003)$}

\def\CCNOCUTLMEBV{$0.161(0.002)$}

\def\CCNOCUTHMRVS{$0.94(0.10)$}

\def\CCNOCUTLMRVS{$1.45(0.08)$}

\def\CCNOCUTHMRVMN{45(90\%)}
\def\CCNOCUTLMRVMN{45(90\%)}
\def\CCNOCUTHMEBVN{27(54\%)}
\def\CCNOCUTLMEBVN{3(6\%)}
\def\CCNOCUTHMRVSN{24(48\%)}
\def\CCNOCUTLMRVSN{16(32\%)}
\def\CCSOMECUTHMRVM{$2.95(0.04)$}

\def\CCSOMECUTLMRVM{$2.17(0.04)$}

\def\CCSOMECUTHMEBV{$0.108(0.001)$}

\def\CCSOMECUTLMEBV{$0.079(0.001)$}

\def\CCSOMECUTHMRVS{$0.66(0.03)$}

\def\CCSOMECUTLMRVS{$0.62(0.13)$}

\def\CCSOMECUTHMRVMN{44(88\%)}
\def\CCSOMECUTLMRVMN{44(88\%)}
\def\CCSOMECUTHMEBVN{35(70\%)}
\def\CCSOMECUTLMEBVN{49(98\%)}
\def\CCSOMECUTHMRVSN{39(78\%)}
\def\CCSOMECUTLMRVSN{33(66\%)}
\def\CCMORECUTHMRVM{$2.91(0.04)$}

\def\CCMORECUTLMRVM{$2.09(0.05)$}

\def\CCMORECUTHMEBV{$0.102(0.001)$}

\def\CCMORECUTLMEBV{$0.076(0.001)$}

\def\CCMORECUTHMRVS{$0.61(0.03)$}

\def\CCMORECUTLMRVS{$0.53(0.12)$}

\def\CCMORECUTHMRVMN{42(84\%)}
\def\CCMORECUTLMRVMN{42(84\%)}
\def\CCMORECUTHMEBVN{18(36\%)}
\def\CCMORECUTLMEBVN{47(94\%)}
\def\CCMORECUTHMRVSN{45(90\%)}
\def\CCMORECUTLMRVSN{36(72\%)}
\def\INTSTEPHMRVM{$3.00(0.03)$}
\def\INTSTEPLMRVM{$2.00(0.03)$}
\def\INTSTEPHMEBV{$0.081(0.001)$}
\def\INTSTEPLMEBV{$0.075(0.001)$}
\def\INTSTEPHMRVS{$0.61(0.02)$}
\def\INTSTEPLMRVS{$0.44(0.04)$}
\def\INTSTEPMSTEP{$0.043(0.002)$}
\def\INTSTEPHMRVMN{48(96\%)}
\def\INTSTEPLMRVMN{47(94\%)}
\def\INTSTEPHMEBVN{48(96\%)}
\def\INTSTEPLMEBVN{48(96\%)}
\def\INTSTEPHMRVSN{48(96\%)}
\def\INTSTEPLMRVSN{49(98\%)}
\def\INTSTEPMSTEPN{49(98\%)}
\def\MsplitZMSFHMRVM{$2.53\pm0.15$}
\def\MsplitZMSFLMRVM{$3.25\pm0.22$}
\def\MsplitZMSFDELRVM{$-0.72\pm0.26$}
\def\MsplitZMSFHMRVS{$0.22 (0.41)$}
\def\MsplitZMSFLMRVS{$0.58\pm0.24$}

\def\MsplitZMSFHMEBV{$0.29\pm0.03$}

\def\MsplitZMSFHMSIG{$0.12\pm0.01$}
\def\MsplitZMSFLMSIG{$0.09\pm0.01$}

\def\MstepZMSFM{$0.038\pm0.022$}
\def\MstepZMSFHMRVM{$2.61\pm0.14$}
\def\MstepZMSFLMRVM{$3.16\pm0.20$}
\def\MstepZMSFDELRVM{$-0.55\pm0.23$}
\def\MstepZMSFHMRVS{$0.27 (0.47)$}
\def\MstepZMSFLMRVS{$0.45\pm0.24$}

\def\MstepZMSFHMEBV{$0.30\pm0.03$}

\def\MstepZMSFHMSIG{$0.11\pm0.01$}
\def\MstepZMSFLMSIG{$0.10\pm0.01$}

\def\MsplitZMSTHMRVM{$2.46\pm0.18$}
\def\MsplitZMSTLMRVM{$3.09\pm0.32$}
\def\MsplitZMSTDELRVM{$-0.64\pm0.35$}
\def\MsplitZMSTHMRVS{$0.28 (0.58)$}
\def\MsplitZMSTLMRVS{$0.85\pm0.34$}

\def\MsplitZMSTHMEBV{$0.23\pm0.02$}
\def\MsplitZMSTLMEBV{$0.26\pm0.03$}

\def\MsplitZMSTHMSIG{$0.10\pm0.01$}
\def\MsplitZMSTLMSIG{$0.07\pm0.02$}

\def\MstepZMSTM{$0.053\pm0.022$}
\def\MstepZMSTHMRVM{$2.57\pm0.20$}
\def\MstepZMSTLMRVM{$2.97\pm0.29$}
\def\MstepZMSTDELRVM{$-0.41\pm0.33$}
\def\MstepZMSTHMRVS{$0.36 (0.70)$}
\def\MstepZMSTLMRVS{$0.76\pm0.35$}

\def\MstepZMSTHMEBV{$0.24\pm0.02$}
\def\MstepZMSTLMEBV{$0.25\pm0.03$}

\def\MstepZMSTHMSIG{$0.10\pm0.01$}
\def\MstepZMSTLMSIG{$0.08\pm0.01$}

\def\ZMSFMgpeak{$-0.019\pm0.045$}
\def\ZMSFMitwenty{$0.139\pm0.039$}
\def\ZMSFMgrpeak{$0.031\pm0.018$}
\def\ZMSTMgpeak{$-0.036\pm0.039$}
\def\ZMSTMitwenty{$0.136\pm0.045$}
\def\ZMSTMgrpeak{$0.042\pm0.016$}

\defcitealias{Grayling24}{G24}

\title{BayeSN and SALT: A Comparison of Dust Inference Across SN~Ia Light-curve Models with DES5YR}

\author[M. Grayling \& B. Popovic]{
M. Grayling$^1$\thanks{Email: mg2102@cam.ac.uk} \&
B. Popovic$^2$\thanks{Email: b.popovic@ip2i.in2p3.fr}
\\
$^1$ Institute of Astronomy and Kavli Institute for Cosmology, Madingley Road, Cambridge CB3 0HA, UK \\
$^2$ Universite Claude Bernard Lyon 1, CNRS, IP2I Lyon / IN2P3, IMR 5822, F-69622 Villeurbanne, France \\
}

\date{Accepted XXX. Received YYY; in original form ZZZ}

\pubyear{2024}

\begin{document}
\label{firstpage}
\pagerange{\pageref{firstpage}--\pageref{lastpage}}
\maketitle

\begin{abstract}
In recent years there has been significant debate around the impact of dust on SNe Ia, a major source of uncertainty in cosmological analyses. We perform the first {validation} of the probabilistic hierarchical SN Ia SED model BayeSN {on} the conventional SALT model, an important test given the history of conflicting conclusions regarding the distributions of host galaxy dust properties between the two. Applying BayeSN to SALT-based simulations, we find that BayeSN is able to accurately recover our simulated inputs {and successfully disentangle differences in dust extinction from an intrinsic mass step. This validates BayeSN as a method to identify the relative contributions of dust and intrinsic differences in explaining the mass step}. When inferring dust parameters with simulated samples including non-Ia contamination, we find that our choice of photometric classifier causes a bias in the inferred dust distribution; this arises because SNe Ia heavily impacted by dust are misclassified as contaminants and excluded. We then apply BayeSN to the sample of SNe from DES5YR to jointly infer host galaxy dust distributions and intrinsic differences on either side of the `mass step' at $10^{10}$ M$\odot$. We find evidence in favour of an intrinsic contribution to the mass step and differing $R_V$ distributions. We also build on recent results supporting an environmental-dependence on the secondary maximum of SNe Ia in $i$-band. Twenty days post-peak, we find an offset in intrinsic $i$-band light curve between each mass bin at a significance in excess of $3\sigma$.
\end{abstract}


%%%%%%%%%%%%%%%%%%%%%%%%%%%%%%%%%%%%%%%%%%%%%%%%%%

%%%%%%%%%%%%%%%%% BODY OF PAPER %%%%%%%%%%%%%%%%%%

\section{Introduction}\label{sec:Intro}

Type Ia supernovae (SNe~Ia) were crucial tools in discovering the accelerating expansion of the universe \citep{Riess98, Perlmutter99}. A possible cause of this expansion, dubbed `dark energy', has been further constrained in subsequent years; \cite{Popovic24a, DES5YR, Brout22} have measured the dark energy equation-of-state $w$ with increasing precision. While the current constraints on $w$ are statistically-dominated, {current and} future surveys such as the Zwicky Transient Facility \citep[ZTF; ][]{Bellm19b, Bellm19a}, the Legacy Survey of Space and Time \citep[LSST; ][]{Ivezic19} and Roman Space Telescope {\citep[Roman;][]{Spergel15, Rose25}} promise orders of magnitude more supernovae, placing a greater need on more precise control of systematic uncertainties. 

To estimate SN~Ia distances with the precision needed to constrain cosmological parameters, a standardisation process is required. Without standardisation, SN~Ia peak-optical absolute magnitudes have a natural scatter of ${\sim}0.40$ mag; however, two empirical relationships have been found that reduce this scatter to ${\sim}0.15$ mag. The first is the Phillips relation, from \cite{Phillips93, Tripp98}, that shows that brighter supernovae have a slower decline rate post-peak; this relation, combined with the discovery that bluer SNe Ia are intrinsically brighter \citep{Riess96}, are crucial to the usage of SNe Ia in the modern day. Most modern cosmological analyses {estimate distances using} the SALT light-curve fitter \cite{Guy07,Guy10, Kenworthy21}.

However, there are some shortcomings to the SALT approach. The process of standardisation within SALT is done through the use of the Tripp formula, a linear relation of magnitude vs. light curve stretch $x_1$ (i.e. the Phillips relation) and {vs.} colour parameter $c$ that roughly corresponds to peak $B-V$ apparent colour. SALT assumes a single colour law for all sources of SN Ia reddening and does not disentangle intrinsic colour variation from dust reddening. These two separate effects are therefore captured by a single linear coefficient for the entire population of SNe Ia; analyses of the host galaxy dust properties of SNe Ia have demonstrated that this approximation is not physical \citep[e.g.][]{Mandel17, BS20, Popovic22, Ward23, Grayling24, Duarte23, Meldorf23, Hallgren25}. 

The current largest systematic uncertainties impacting cosmology with SNe Ia are astrophysical \citep{Popovic24a, DES5YR}, and decreasing these uncertainties will require further understanding of the complex astrophysical processes that drive SNe Ia {and their environments}. \cite{Dhawan24} recently explored how a variety of effects, including evolution of SN Ia host galaxy dust properties with redshift, would lead to offsets from a fiducial $\Lambda$CDM cosmology {comparable to that seen in \citet{DESI24, DESI25} when combining SN constraints with baryonic acoustic oscillation (BAO) results from the Dark Energy Spectroscopic Instrument \citep[DESI; ][]{DESI22}}. In the past decade, new discoveries have improved our understanding of how SNe Ia relate to their host galaxies. The `mass step', the observation that SNe Ia in higher mass galaxies are brighter than their low mass counterparts post-standardisation, was discovered by \cite{Sullivan10, Kelly10, Lampeitl10}. Since the initial discovery of the mass step, it has been consistently observed in a variety of independent optical samples \citep{Childress14, Betoule14, Jones18, Smith20, Kelsey21, Kelsey23}. This observation has been further extended to other host-galaxy environmental properties, including star formation rate (SFR), specific star formation rate (sSFR), and host-galaxy colour \citep{Rigault18, Briday21, Kelsey21, Kelsey23, Roman18}. {A number of studies have considered properties of the local environments of SNe Ia rather than those derived from the entire host galaxy \citep[e.g.][]{Rigault13, Rigault15, Rigault20, Jones15, Moreno-Raya16b, Moreno-Raya16a, Jones18, Kim18, Roman18, Kim19, Rose19, Kelsey21}}, sometimes finding that this leads to a stronger environmental dependence.

From the SALT side, \cite{BS20} and later \cite{Popovic22} have had success in attributing the root cause of the mass step to dust. These papers showed that a single population of intrinsic SN~Ia colours, reddened and dimmed by dust from the host-galaxy, is able to recreate observed features of SN~Ia data. Specifically, these works postulated that the mass step is a result of a difference in the distribution of SN Ia line-of-sight $R_V$ values between high- and low-mass galaxies, with a difference in the mean $R_V$ value $\mu_{R_V}$ in each environment of $\Delta\mu_{R_V}\sim1.2-1.6$.. This dust-based scatter model was used in the \cite{Brout22, DES5YR, Popovic24a} cosmological analyses. 

However, these works assumed a mass step entirely caused by dust and did not consider the possibility of an intrinsic component, {which has been suggested by a number of other works. For example, \citet{Gonzalez-Gaitan21} finds an intrinsic contribution to the mass step alongside a difference in $R_V$ values between high- and low-mass galaxies and speculates  that this could result from a difference in intrinsic colour. \citet{Duarte23} found that host galaxy dust differences cannot fully explain the mass step through analysis of SN host galaxy attenuation laws in DES. \citet{Wiseman23} finds that a model with a galaxy age-varying $R_V$ and no intrinsic luminosity difference replicates the observed SN population well, although does not rule out that intrinsic differences may play a role in the mass step. Recent analyses of DES \citep{DES5YR} and ZTF \citep{Ginolin24b} have also shown that a mass step persists even when considering only the bluest SNe with minimal impact from dust, providing further evidence for an intrinsic contribution to the mass step. \citet{Wiseman23} finds a model with a galaxy age-varying $R_V$ and no intrinsic luminosity difference can reproduce the observed SN population but does not rule out an intrinsic contribution to the mass step.} As shown by e.g. \citet{Thorp21, TM22, Grayling24, Popovic24c}, attributing all sources of magnitude variation in SNe Ia to dust may overestimate the difference in $R_V$ values between SNe Ia in high- and low-mass galaxies, and does not properly account for sources of scatter in the data that do not correlate with the SN Ia colour parameter. \cite{Grayling24}, hereafter \citetalias{Grayling24}, finds evidence in favour of an intrinsic mass step ${\sim}0.05$ mag, inferred even when allowing for different host galaxy dust distributions either side of the mass step; they also find tentative evidence for a difference in peak intrinsic colour between SNe Ia on either side of the mass step. Moreover, \citetalias{Grayling24} found a 4.5$\sigma$ difference in intrinsic absolute magnitude in $i$-band 20 days post-peak between SNe Ia in high and low-mass galaxies, providing further evidence in favour of intrinsic differences between SNe Ia in different environments. {These intrinsic differences are inferred alongside much smaller values of $\Delta\mu_{R_V}$ than SALT-based analyses.}

Given the differing conclusions regarding dust { -- specifically, differences in $R_V$ distributions in different environments --} from SALT and BayeSN analyses, it is important to understand exactly where these differences arise and try to reconcile these results. In this work, we perform the first {comparison} between the two by applying BayeSN for inference of dust properties on SALT-based simulations similar to those used in the DES5YR cosmology analysis. Following from this, we apply BayeSN to a sample of SNe Ia from the DES5YR sample to comment on the relative contributions of dust and intrinsic effects in explaining the mass step. We also compare these results on the real DES5YR sample to those from SALT-based analyses.

The layout of this paper is as follows. Section \ref{sec:LCFitters} gives an overview of the BayeSN and SALT SN Ia SED models, as well as the simulation framework. Section \ref{sec:Data} details the sample of SNe Ia from DES5YR we use in this analysis, and the methodology is detailed in Section \ref{sec:Methodology}. Section \ref{sec:simreco} presents recovery of simulated parameters from applying BayeSN to SALT-based simulations. Our results from applying BayeSN to the DES5YR sample are presented in Section \ref{sec:Results:subsec:dust}, followed by discussion and conclusions in Section \ref{sec:Conclusions}.

\section{SED Models: BayeSN and SALT}\label{sec:LCFitters}

\subsection{BayeSN}\label{sec:LCFitters:subsec:BayeSN}

BayeSN is a probabilistic SED model for SNe Ia, which uses hierarchical Bayesian modelling to jointly infer individual and population-level properties of SNe Ia \citep[e.g.][]{Mandel09, Mandel11, Mandel17}. This enables inference of SN Ia population parameters such as host galaxy dust distributions, while also creating a Spectral Energy Distribution (SED) model that can be used for light curve fitting. While a full description is available in \cite{Mandel22, Grayling24}, we provide a broad overview of the model here. The full BayeSN time- and wavelength-dependent SED model is given by:  
\begin{equation}
\label{bayesn_equation}
\begin{aligned}
    -2.5\log_{10}[S_s(t,\lambda_r)/S_0(t,\lambda_r)] = M_0 + W_0(t,\lambda_r) \ + \ \\ \delta M^s+\theta^s_1W_1(t,\lambda_r) + \epsilon^s(t,\lambda_r)+A^s_V\xi\big(\lambda_r;R^{(s)}_V\big).
\end{aligned}
\end{equation}
BayeSN uses the optical-NIR SN~Ia SED template from \cite{Hsiao11} as the base SED, defined as $S_0(t, \lambda_r)$, with an arbitrary scaling factor $M_0$ of $-19.5$. The rest-frame wavelength and rest-frame phase relative to maximum flux are given by $\lambda_r$ and $t$ respectively. Latent variables which are different for each SN are denoted with $^s$; the other parameters are global hyperparameters shared across the whole population. The different components which make up the model are as follows:
\begin{itemize}
    \item $W_0(t, \lambda_r)$ warps and normalises the zeroth-order Hsiao SED template to establish the mean intrinsic SED for the population. $W_1(t, \lambda_r)$ is a functional principal component that describes the first-order mode of variation of the intrinsic SED across the population.
    \item $\theta_1^s$ is a principal component coefficient which quantifies the effect of the $W_1$ functional principal component (FPC) for each supernova. In concert with $W_1(t, \lambda_r)$, this accounts for the `broader-brighter' relationship observed across SNe Ia \cite{Phillips93}. $\theta_1$ is defined to follow a normal distribution such that $\theta_1\sim\mathcal{N}(0,1)$.
    \item $\delta M^s$ is an achromatic, time-independent offset for each SN; this parameter is drawn from a normal distribution with $\delta M^s \sim N(0, \sigma_0^2)$, where $\sigma_0$ is a hyperparameter that represents the intrinsic achromatic scatter seen across the SN Ia population.
    \item $\epsilon^s(t,\lambda_r)$ is a time- and wavelength-dependent cubic spline function that captures residual intrinsic colour variations, defined by a matrix of knots $\mathbf{E}^s$. This $\mathbf{E}^s$ parameter is drawn from a multivariate Gaussian $\mathbf{e}^s \sim N(0, \mathbf{\Sigma}_\epsilon$), where the $\mathbf{E}^s$ matrix has been vectorised as $\mathbf{e}^s$. The associated covariance matrix $\mathbf{\Sigma}_\epsilon$ is a hyperparameter that details the distribution of residual intrinsic scatter at the population level.
    \item $A_V^s$ and $R_V^s$ are host galaxy extinction parameters for each SN; $A_V$ is the total extinction, compared to the selective-extinction-ratio $R_V$ that parameterises the \cite{Fitzpatrick99} dust extinction law, here $\xi\big(\lambda_r;R^s_V\big)$. $A_V$ is assumed to follow an exponential prior with a population mean/scale parameter $\tau_A$; which is to say, $A_V \sim \text{Exponential}(\tau_A)$. $R_V$ follows a normal distribution truncated at $R_V = 1.2$, $R_V^s \sim TN(\mu_{R_V}, \sigma_{R_V}^2, 1.2, \infty)$\footnote{As discussed in Section 3.2.3 of \citetalias{Grayling24}, $\mu_{R_V}$ and $\sigma^2_{R_V}$ refer to the mean and variance of the distribution \textit{before} truncation. The population mean and variance of the truncated distribution, $\mathbb{E}[ R_V ]$ and $\text{Var}[R_V]$, are not necessarily the same as $\mu_{R_V}$ and $\sigma^2_{R_V}$. Within this work, however, we find that the impact of truncation is small so opt to focus solely on $\mu_{R_V}$ and $\sigma^2_{R_V}$ values.}, with the $R_V = 1.2$ limit motivated by the Rayleigh scattering limit \citep{Draine03}, though we modify this limit for our SALT simulations (see Section \ref{sec:Methodology:subsec:simreco}).
\end{itemize}

The resulting rest-frame, host galaxy extinguished SED $S_s(t, \lambda_r)$ is then corrected for Milky Way dust extinction using dust maps from \cite{Schlafly11} with an assumed $R_V = 3.1$ following \cite{Fitzpatrick99}, before being scaled by the distance modulus $\mu^s$ and redshifted. This final SED can be integrated through filters to produce synthetic photometry that is used in tandem with observed photometry to calculate likelihoods.

\subsubsection{BayeSN Treatment of the Mass Step}
\label{sec:LCFitters:subsec:BayeSNmassstep}

As described in \citetalias{Grayling24}, BayeSN uses a flexible treatment to allow for a mass step caused by dust, some other intrinsic effect or some combination thereof. \cite{Thorp21, TM22} first applied BayeSN to an investigation of the mass step by jointly inferring host galaxy dust properties of SNe Ia in different environments jointly with an intrinsic, achromatic mass step parameter. This approach was extended in \citetalias{Grayling24} to allow for time- and wavelength-dependent intrinsic differences between SNe Ia in different environments, as discussed in Section 3.3.2 of that work.

Here, we build upon the model developed in \citetalias{Grayling24}. In that work, $W_0(t, \lambda_r)$ and $W_1(t, \lambda_r)$ were fixed to the values inferred in \cite{Thorp21} based on a sample of 157 SNe Ia in the Foundation SN~Ia sample \citep{Foley18}. Here, we jointly infer $W_0(t, \lambda_r)$ and $W_1(t, \lambda_r)$ with all other hyperparameters. Adding this extra flexibility to the model was necessary to allow BayeSN to infer an intrinsic SED which matched the SALT-based simulations it has been applied to.

In all instances, when considering separate populations of SNe Ia binned by stellar mass we allow for each population to have its own population of host galaxy dust properties. For example, the distribution on $R_V^s$ is:

\begin{equation}
    R_V^s \sim
    \begin{cases}
        TN(\mu_{R_V,\text{HM}}, \sigma_{R,\text{HM}}^2, 1.2, \infty), & \text{ if } M_*^s >  M_\text{split} \\
        TN(\mu_{R_V,\text{LM}}, \sigma_{R,\text{LM}}^2, 1.2, \infty), & \text{ if } M_*^s <  M_\text{split} \\
    \end{cases}
\end{equation}
where $M_*^s$ is the host stellar mass of each SN $s$ and $M_\text{split}$ is some reference stellar mass at which the split point is located, set at $ 10^{10} M_\odot$ in this work. We also allow for separate $A_V$ distributions for high- and low-mass galaxies described by different $\tau_A$ parameters. 

We consider two possible parameterisations of an intrinsic mass step in this work. In the first, we consider a mass step parameter $\Delta M_0$ that acts as an achromatic, constant magnitude offset between the intrinsic SEDs of SNe in high- and low-mass galaxies. This is implemented as a shift on the mean of the achromatic scatter distribution such that

\begin{equation}
    \delta M^s \sim
    \begin{cases}
        N(0, \sigma_{0,\text{HM}}^2), & \text{ if } M_*^s >  M_\text{split} \\
        N(\Delta M_0, \sigma_{0,\text{LM}}^2), & \text{ if } M_*^s <  M_\text{split}. \\
    \end{cases}
\end{equation}

For the second intrinsic mass step parameterisation, we infer a separate $W_0(t, \lambda_r)$ for SNe Ia in each mass bin. This allows for time- and wavelength-dependent differences between each mass bin. We refer to the first case as an `intrinsic mag difference', and the second case as an `intrinsic SED difference'.

\subsubsection{Conditioning on Distance}

As discussed in \citet{TM22, Grayling24}, two approaches can be taken regarding distance in this type of hierarchical model. It is possible either to leave photometric distance as a free parameter and marginalise over the distance to each SN $\mu^s$ when inferring dust hyperparameters, or to assume a cosmology and condition on distance based on redshifts. While the former approach provides cosmology-independent constraints, it is more suited to optical plus NIR analyses with more colour information available. With optical-only analyses conditioning on redshift-based distances provides far stronger dust constraints. We follow the approach outlined in Section 3.2.2 of \citetalias{Grayling24} in conditioning on an assumed cosmology.

For this work, in our analysis of the DES5YR sample we assume a flat $\Lambda$CDM cosmology with $\Omega_m=0.28$ and $H_0$~$=$~73.24~km~s$^{-1}$~Mpc$^{-1}$ for consistency with previous BayeSN analyses. For our inference on SALT-based simulations, however, we assume a flat $\Lambda$CDM cosmology with $\Omega_m=0.311$ and $H_0 = 70$ km s$^{-1}$ in order to match the cosmology assumed when creating the simulations.

\subsection{Simulations with SALT }\label{sec:LCFitter:subsec:Sims}

Here, we focus on the compatibility of BayeSN and SALT. To-date, there has been no cross-comparison between BayeSN and SALT-based approaches to dust inference. Given the history of conflicting results between the dust parameters derived from BayeSN \citep{Mandel22,Thorp21,Grayling24} and those from SALT \citep{BS20, Wiseman22, Popovic22}, in particular Dust2Dust \citep{Popovic22}, we focus on the ability of BayeSN to recover dust parameters from SALT-based simulations of SN~Ia light-curves, and additionally, contaminated samples. The self-consistency of Dust2Dust is reviewed in \cite{Popovic22}. 

To simulate SNe Ia populations with dust, we use {\sc snana} \citep{SNANA, Kessler19}. This generates a base SED, which is integrated through filter transmission functions  to create model fluxes that then have noise and detection criteria applied to provide simulated photometry measurements.

We aim to {validate the perfomance of BayeSN and examine differences between the SALT and BayeSN treatments of dust}, and therefore our base source model is the SALT model \citep{Guy07}, updated {to SALT3} by \cite{Kenworthy21}. Compared to BayeSN, SALT models the flux as 
\begin{align} 
\label{saltmodel}
\begin{split}
F(\rm{SN}, p, \lambda) = x_{0}^s &\times\left[M_{0}(p, \lambda)+x_{1}^s M_{1}(p, \lambda)+\ldots\right] \\
&\times \exp [c^s C L(\lambda)],
\end{split}
\end{align}
which results in a similar SED-model: $M_0$\footnote{{BayeSN uses the Hsiao 07 template as its base for $W_0$, SALT does not use a pre-existing template as the basis for $M_0$.}} and $M_1$ roughly correspond to $W_0$ and $W_1$ in BayeSN, modelling the average SED and the first-order mode of variation to the SED respectively. The colour law $CL(\lambda)$ captures the colour-related variation in the SED across the population, combining intrinsic variation with the effect of host galaxy dust extinction which are respectively modelled by $\mathbf{\Sigma_\epsilon}$ and $\xi$ in BayeSN. SALT $x_1^s$ finds its BayeSN analogue in $\theta^s_1$, while $c$ is roughly comparable though not analogous to $A^s_V$. In SALT, the overall amplitude of the light-curve at peak brightness, $x_0^s$, has no direct correspondent in BayeSN, which infers the distance jointly with the other parameters. $x_1^s$, $c^s$, and $x_0^s$ are all determined for each supernova, whereas $M_0$ and $M_1$ are population-level SEDs that are determined during the SALT training process.

Within the SALT framework, distances are defined with the Tripp estimator \citep{Tripp98}:
\begin{equation}
\label{eq:tripp}
    \mu = m_B + \alpha_{\rm SALT} x_1 - \beta_{\rm SALT} c - M_0
\end{equation}
where $m_B = -2.5 \textrm{log}_{10}(x_0^s)$, and $x_1$ and $c$ are defined previously, with the $^s$ superscript dropped for convenience. $M_0$ in the Tripp formulation is the absolute magnitude of an SN~Ia with $x_1 = c = 0$, and $\alpha_{\rm SALT}$ and $\beta_{\rm SALT}$ are population-level nuisance parameters.

The dust-modeling framework introduced in \cite{Mandel17, BS20, Popovic22} makes some changes to the SALT and Tripp formulations. The distribution of SN~Ia colours is comprised of a combination of a intrinsic, dust-free distribution - $c_{\rm int}$ - and a reddening component due to dust ($E_{\rm dust}$). The $c$ component in Equation \ref{eq:tripp} is then 
\begin{equation}
    c = c_{\rm int} + E_{\rm dust}.
\label{eq:cobs}
\end{equation}
This $E_{\rm dust}$ component is related to the total extinction $A_V$ as $A_V = R_V \times E_{\rm Dust}$, such that $E_{\rm Dust} = E(B-V)$ and $R_V$ is the total-to-selective extinction ratio.

The dust scatter models then dim the SALT SED according to \cite{Fitzpatrick99} with a given $R_V$; this can be understood to change the overall observed brightness as 
\begin{equation}
    \Delta m_B = \beta_{\rm SN}c_{\rm int} + (R_V+1)E_{\rm dust}
    \label{eq:deltamb}
\end{equation}
where $\beta_{\rm SN}$ is a colour-luminosity relationship intrinsic to the supernova. This process is elucidated further in \cite{BS20} and \cite{Popovic22}. It should be noted that while $c_{\rm int}$ is treated as a purely intrinsic colour term, when forward simulating SNe the effect of $c_{\rm int}$ on the underlying SED is still the default SALT colour law $CL(\lambda)$, itself a combination of dust and intrinsic colour variation. This means that in reality $c_{\rm int}$ will have some dust-related contribution, although this will correspond to a single $R_V$ for all SNe and is likely to be small.

The dust-based SALT simulation framework we use here generates light curves with a SALT $x_1$ value; we use the $x_1$ distribution from \cite{DES5YR} for this work. 

\subsection{Differences between BayeSN and SALT: Dust}
\label{sec:LCFitters:subsec:modeldiff}

As part of this work, we apply BayeSN to infer dust properties on SALT-based simulations of the DES5YR sample. Our objective is to test for consistency between the two models and validate BayeSN's ability to generalise to dust inference using simulations created using a different model. However, it is important to emphasise that we do not necessarily expect completely unbiased parameter recovery between the SALT simulations and BayeSN dust inference. This is because there are a number of fundamental differences between the two, as discussed below. This list is not exhaustive, but gives a sense of some of the key differences:

\begin{itemize}
    \item The SALT simulations include a treatment of weak lensing, which magnifies the `observed' light curves, however BayeSN does not incorporate the effect of weak lensing. At the low redshifts of $z\leq0.36$ we consider in this work this effect will be small, although not entirely negligible.
    \item BayeSN assumes a constant, achromatic scatter $\delta M^s$ around the distance-redshift relation which follows a Gaussian distribution, in addition to a chromatic scatter term.
    {SALT does not explicitly include a grey scatter term equivalent to $\delta M^s$, though conventional cosmology analyses with SALT will approximate the grey scatter as a Gaussian.}
    \item As discussed in Section \ref{sec:LCFitter:subsec:Sims}, the $c_{\rm int}$ parameter used in the SALT sims will include a small dust contribution from the SALT colour law in addition to the \cite{Fitzpatrick99} extinction law used. BayeSN, on the other hand, only considers a dust contribution from the \cite{Fitzpatrick99} extinction law.
\end{itemize}

Nevertheless, while some differences may be expected we hope that by jointly inferring an intrinsic SED for SNe Ia simultaneously with dust, BayeSN has enough flexibility to recover the simulated dust parameters. We do hope for a good level of agreement between the two models.

\section{Data}\label{sec:Data}

We make use of the Dark Energy Survey \lq 5-year\rq\ data release (DES-SN, \citep{Flaugher15, Sanchez24}. The DES observing program ran for 5 years, observing likely SNe Ia from $0.1<z<1.13$ with $griz$ bands. The host galaxy information for the DES supernovae is taken from co-added images \citep{Wiseman20}, with environmental properties such as stellar mass and colour fit from the spectral energy distributions by \cite{Smith20, Kelsey23}. A spectroscopic follow-up program by the Australian DES survey, OzDES \citep{Childress17, Lidman20}, provided spectroscopic redshifts of the host galaxies, which was paired with the state-of-the-art photometric classification program \texttt{SuperNNova} from \cite{Moller19}. The resulting DES-SN sample is the largest single-telescope sample to-date, comprised of 1500 likely SNe Ia. A more comprehensive review of the SNe Ia used in the DES 5-year analysis can be found in \cite{DES5YR, Moller24}.

\subsection{Establishing the BayeDESN Sample}\label{sec:Data:subsec:sample}

We begin by establishing a sample of DES5YR SNe Ia to use for population inference with BayeSN. Unlike Dust2Dust, based on SALT, BayeSN is a hierarchical Bayesian model with a joint parameter space of population-level hyperparameters and individual object latent parameters. This work is the first time that BayeSN has been applied to a photometrically-classified sample, and the model does not currently incorporate a treatment for contamination from core-collapse SNe. The presence of contaminants, or indeed poor quality data (see discussion in Appendix \ref{sec:Appendix}), has the potential to bias the inferred population-level parameter values. Therefore, we apply a number of cuts to create our final sample.

Our objective with this sample is to include as many objects as possible while making a minimal set of reasonable quality cuts. Cosmology samples using SALT typically follow a set of well established cuts on e.g. $x_1$, $c$ \textit{after the SALT fitting process}; these selection cuts typically discard reddened SNe with $c > 0.3$, which are the most affected by dust and give the strongest constraint on its properties. Given that we are exploring an alternative model, we aim to avoid selecting a sample based on SALT. The starting point for our sample, however, is the complete set of light curves from the DES data release that passed SALT fits. This has two motivations; firstly, it reduces the number of transient light curves observed by DES from 19,702 to 3,621. Secondly, the requirement of passing SALT fits merely requires that the SALT fit converged, rather than any stringent quality cuts. In effect, this starting point will give us a sample of reasonable light curves that resemble SNe, without any strong selection on their properties.

From this SALT-converged set, we use the \texttt{SuperNNova}-assigned probability of being an SN~Ia $P_{\rm Ia} > 0.5$, to eliminate likely non-Ia supernovae. We test the validity of this assumption in Section \ref{sec:simreco:subsec:CC}. In addition, we apply a very loose cut on Hubble residual to remove remaining likely contaminants, requiring $|\Delta\mu_\text{res}|<1.0$. 

Capturing selection effects within a hierarchical Bayesian model is a challenging problem, and BayeSN does not incorporate such a treatment at present. Therefore, we create a volume-limited sample to mitigate for survey selection effects, applying a redshift cut of $z \leq 0.36$ following \cite{Popovic24c}.

We place a number of additional quality cuts, including:
\begin{itemize}
    \item Requiring an uncertainty on time-of-maximum, as estimated by fitting the light curves using the BayeSN model trained in \cite{Thorp21}, of less than 2 rest-frame days; $\sigma_{\text{t,max}} < 2$.
    \item Requiring a data point within 5 rest-frame days of the estimated time-of-maximum. We impose this cut, rather than requiring data both before and after peak, to include a number of well-constrained light curves with data very close to peak but none before.
    \item Requiring a host galaxy stellar mass present in the public DES5YR data release.
    \item Requiring data in at least two bands usable by BayeSN. Please note that in this work we have used the same wavelength range for the BayeSN model as in \cite{Thorp21, Grayling24}, with the lower bound of the model at 3500 \AA. SALT, meanwhile, is defined down to a lower wavelength cut-off. There are a small number of supernovae for which $g$-band data has been excluded in our BayeSN fits but would have been used in the SALT analysis. This is why there are a number of SNe which passed the SALT band coverage requirements which are nonetheless excluded in this analysis. Improving the coverage of BayeSN into bluer wavelengths is a goal for future work.
    \item We cut three supernovae which have non-standard light curves which were found to stop the model converging. In two cases, light curves contained multiple, discrepant high signal-to-noise photometric data points in the same band at very similar epochs, which would pose a great challenge for any model. For the third case, the light curve showed an extra peak in the light curve not expected for typical SNe Ia. These examples are shown in Appendix \ref{sec:Appendix}.
\end{itemize}

{In short, these cuts give us the 393 SNe that define the BayeDESN sample}. We additionally consider an alternative sample, the `Cosmo' sample, which follows these requirements but additionally includes only SNe in the final DES5YR cosmology sample. The specific cuts applied to these samples, and the number of SNe they respectively cut, are detailed in Table \ref{tab:cuts}.

\begin{table}
\centering
\begin{tabular}{c|cc}
\textbf{Cut} & \multicolumn{2}{c}{\textbf{Total SNe}} \\
 & (Cosmo) & (BayeDESN) 
\\
\hline
SALT3 fit converged and $z>0.025$ &  \multicolumn{2}{c}{3621} \\
$P_{\rm Ia} > 0.5$ & \multicolumn{2}{c}{2201} \\
$z \leq 0.36$ Cut &  \multicolumn{2}{c}{423} \\
In DES5YR cosmo sample & 343 & 423 \\
$|\Delta\mu_\text{res}|$ < 1.0 & 342 & 417 \\
$\sigma_{t\text{,max}} < 2$ days & 342 & 411 \\
Data within 5 rest-frame days of peak brightness & 342 & 407 \\
Missing host mass & 342 & 401 \\
Only one usable band & 338 & 396 \\
Quality cuts & 336 & 393 \\
\hline
\textbf{Final} & \textbf{336} & \textbf{393} \\
\end{tabular}
    \caption{Quality Cuts Establishing the `Cosmo' and BayeDESN Sample. We remove 3 total LCs based on eye tests.}
    \label{tab:cuts}
\end{table}

\section{Methodology}\label{sec:Methodology}

{The focus of this work is to apply BayeSN to infer dust properties on SALT-based simulations, given the history of differing results concerning dust between the two. This will test BayeSN's ability to accurately infer dust properties and potentially illustrate the reason for previous differences in results.} We perform a suite of tests with simulated data, as detailed in sub-sections \ref{sec:Methodology:subsec:simreco} and \ref{sec:Methodology:subsec:CC}. {We present the results of these tests in Section \ref{sec:simreco}} then, in Section \ref{sec:Results:subsec:dust}, we apply BayeSN to the real BayeDESN sample, to infer differences in dust and intrinsic properties on either side of the mass step.

\subsection{Recovery of Simulated Parameters}\label{sec:Methodology:subsec:simreco}

Following Section \ref{sec:LCFitter:subsec:Sims}, we simulate 50 simulacra of the DES-5YR dataset, varying supernova and dust properties according to Table \ref{tab:inputs}. These simulations are then fit using the BayeSN model for hierarchical inference. We infer all global parameters including $W_0, W_1, \Sigma_{\epsilon}$ as well as the mean and standard deviation of the $R_V$ population, the population mean extinction $\tau_A$, and an achromatic intrinsic scatter $\sigma_0$; this allows the BayeSN model the flexibility to capture the variation in the SALT SED model while simultaneously inferring dust properties. In the cases where our simulated dust properties vary with host galaxy mass (see Table \ref{tab:inputs}), we similarly split our hyperparameters to match the simulated `step' at $10^{10} M_{*}$, where our dust distributions change. In these cases, we apply the `intrinsic SED difference' model discussed in Section \ref{sec:LCFitters:subsec:BayeSNmassstep} {to apply the most general model possible, with the exception of \ref{sec:Results:subsec:dust:subsubsec:intdiff} where we attempt to constrain a simulated intrinsic mass step and therefore apply the `intrinsic mag difference' model to directly infer a mass step parameter.}

\begin{table*}
    \centering
    \begin{tabular}{c|ccc}
        Systematic Test & $R_V$ & $E(B-V)$ & $\beta_{\rm SN}$ \\
        \hline
        Single $R_V$, $E(B-V)$ & $\mathcal{N}(3, 0.5)$ & $\exp(0.080)$ & 1 \\ 
        Double $R_V$, Single $E(B-V)$ & $\mathcal{N}(2, 0.5), \mathcal{N}(3, 0.5)$ & $\exp(0.080)$ & 1 \\ 
        Double $R_V$, $E(B-V)$ & $\mathcal{N}(2, 0.5), \mathcal{N}(3, 0.5)$ & $\exp(0.125)$, $\exp(0.080)$ & 1 \\ 
        Multi-$\beta$ & $\mathcal{N}(2, 0.5), \mathcal{N}(3, 0.5)$ & $\exp(0.125)$, $\exp(0.080)$ & $\mathcal{N}(2, 0.2)$
    \end{tabular}
    \caption{A summary of the distributions of dust parameters and $\beta_{SN}$ used in each systematic test.}
    \label{tab:inputs}
\end{table*}

When applying BayeSN to these simulations, we vary the lower bound on the truncated distribution of $R_V$ values from 1.2 to 0.4. While the lower limit of 1.2 is motivated by the Rayleigh scattering limit, the SALT simulations we apply the model to use a lower bound of 0.4. Varying the lower bound for BayeSN ensures that the full range of simulated values is allowed when inferring dust properties on these simulations. 

We use a uniform hyperprior $U(0.4, 6)$ on the mean of the $R_V$ distribution, $\mu_{R_V}$.  For the standard deviation of the $R_V$ distribution, we use a half-normal hyperprior with a scale factor of 2: $\sigma_R \sim \text{Half-}N(0, 2^2)$. This prior is deliberately wide to encompass the full range of $\sigma_R$ values which have previously been proposed in the literature. For $\sigma_0$ and $\tau_A$, we use half-Cauchy hyperpriors with scale factors of 0.1 mag and 1 mag respectively, following \cite{Grayling24, Mandel22}. 

\subsection{Impact of non-Ia Contamination}\label{sec:Methodology:subsec:CC}

Here, we aim to test the impact of non-Ia contamination on dust inference using BayeSN. To do so, we simulate the Double $R_V$, Double $E(B-V)$ test with the addition of templates of non-Ia SNe from \cite{Vincenzi20}. We include four classes of non-Ia, following \cite{DES5YR}: SN Iax and SN~Ia 91bg from \cite{Kessler19} following revisions from \cite{Vincenzi19}, and core collapse SED templates from \cite{Vincenzi19} with rates from \cite{Strolger15} and \cite{Shivvers17} with the modified \cite{Li11} luminosity functions from \cite{Vincenzi20}. A full description of the non-Ia templates is available in \cite{Vincenzi20}. We classify our simulated SN light-curves using the \texttt{SuperNNova} classifier from \cite{Moller19}, taking the trained \texttt{SuperNNova} model directly from the DES 5-year analysis, where a further review is available. 

With the simulated SNe (non-Ia and Ia), we perform a BayeSN population inference three times; first with no \texttt{SuperNNova} probability cut, then again with a \texttt{SuperNNova} $P_{\rm Ia} >0.5$ and a \texttt{SuperNNova} $P_{\rm Ia} >0.9$ cut placed on our simulated sample, in order to compare the impact of varying levels of contamination on our dust inference.

\section{Simulation Recovery}\label{sec:simreco}

In this section we present the results of BayeSN dust inference applied to SALT simulations. Section \ref{sec:simreco:subsec:metrics} details the statistics we have used to assess the consistency of the BayeSN posteriors with the true simulated parameters, before presenting the results in Sections \ref{sec:simreco:subsec:puresimresults} and \ref{sec:simreco:subsec:CC}.

\subsection{Evaluation Metrics}
\label{sec:simreco:subsec:metrics}

For each simulation variant outlined in Section \ref{sec:Methodology:subsec:simreco}, we apply BayeSN to obtain posterior distributions on dust hyperparameters $\mu_{R_V}$, $\sigma_R$ and $\tau_E$ for all 50 simulacra. Our goal in this analysis is to test the consistency between the BayeSN posteriors and the SALT-simulated dust parameters. We calculate a number of metrics to explore this.

We will denote each hyperparameter generically as $\phi$, where $\phi\in\{\mu_{R_V}, \sigma_R, \tau_A\}$; the true simulated value is referred to as the `Input Value' in Tables \ref{tab:1RV1EBVReco} to \ref{tab:FULLReco}. For each simulation variant we have 50 sets of posteriors $\phi_i$, each with a posterior mean $\hat{\phi_i}$ and variance $\sigma_{\phi,i}^2$, where the index $i$ denotes the individual simulacrum of the simulation. We are interested in the overall BayeSN-estimated dust hyperparameters across all 50 simulacra, for which we calculate a number of different metrics. The first is the overall mean of the individual $\hat{\phi_i}$ values (`Estimated Value' in our tables), which we denote as $\Bar{\phi}$ and calculate using

\begin{equation}
    \Bar{\phi} = \frac{\sum_i\sigma_{\phi,i}^{-2}\hat{\phi_i}}{\sum_{i}\sigma_{\phi,i}^{-2}}.
\end{equation}

The mean $\Bar{\phi}$ is weighted by $\sigma_{\phi_i}$ so that the relative confidence of each individual $\hat{\phi_i}$ value is reflected in the overall summary statistic. We also present the standard error on this mean, $\sigma_{\Bar{\phi}}$ (Std. Err.), defined as $\sigma_{\Bar{\phi}}=\sqrt{\text{Var}(\hat{\phi_i})/50}$.

For each hyperparameter, we define a coverage metric $N_{95}$\footnote{We considered additional coverage metrics $\sigma_{\hat{\phi}_i}$, the standard deviation across posterior means $\hat{\phi_i}$ and $\Bar{\sigma}_\phi$, the mean posterior uncertainty across all 50 simulacra. Where the posteriors have good coverage of the truth, we would expect these two values to be identical or very similar. In practice, this was what we generally found. In cases where they differed, this lack of posterior coverage of the truth was reflected by $N_{95}$. We therefore opt to focus solely on $N_{95}$ when assessing posterior coverage for brevity.}; this corresponds to the number of our 50 simulated simulacra where the true simulated parameter falls within the 95 per cent credible region of the posterior $\phi_i$. When applied to simulations of the same model, we would expect this value to be ${\sim}47-48$, around 95 per cent of the simulations. As we have applied BayeSN to SALT-based simulations, however, this will not necessarily be the case.

\subsection{BayeSN Dust Estimates}
\label{sec:simreco:subsec:puresimresults}

The true simulated values input into the Dust2Dust simulations and our BayeSN posterior metrics across all 50 simulacra of our simulations are detailed in Tables \ref{tab:1RV1EBVReco} through \ref{tab:INTReco}.

Overall, BayeSN demonstrates an excellent ability to accurately recover simulated dust properties even with an entirely distinct SN model as the basis for simulations. As discussed in Section \ref{sec:LCFitters:subsec:modeldiff}, fundamental differences between the two models mean that we hope for good agreement but do not necessarily expect unbiased parameter recovery. Nevertheless, these results show that the simulated parameter values are recovered near-perfectly.

Firstly, considering $\mu_{R_V}$ across all simulated variations the BayeSN estimates are in very close agreement with the truth, within $1\sigma$ in many cases. The most notable discrepancy is $\Bar{\mu}_R=2.12(0.02)$ for the $\mu_{R_V}=2.0$ high-mass bin in the case of the double $R_V$, $E(B-V)$ simulation with the Gaussian $\beta_{\rm SN}$ distribution, shown in Table \ref{tab:FULLReco}. The Gaussian $\beta_{\rm SN}$ distribution has no equivalent in BayeSN, and was chosen to test potentially diverging choices in how the two models treat dust; our expectation was that this would be the simulated variant where BayeSN would perform the worst. A discrepancy of ${\sim}0.1$ in $\mu_{R_V}$ is small compared to the statistical uncertainty on real inferred $\mu_{R_V}$ values; this result could not explain previous disagreements between the findings of BayeSN and Dust2Dust. Additionally, the estimate of $\Bar{\mu}_R$ in the low-mass bin for this simulation exactly matches the truth of 3.0, showing that this specific model mismatch does not cause particular issues for BayeSN. Our posteriors on $\mu_{R_V}$ are relatively well-calibrated considering the inherent model-mismatch in this analysis; $N_{95}$ is close to 95 per cent across all of our simulation variants for $\mu_{R_V}$.

BayeSN estimates of $\tau_E$ are also excellent, in close agreement with the truth. It is worth noting that BayeSN does not actually fit for $\tau_E$, placing an exponential distribution on $A_V^s$ rather than $E(B-V)^s$ as in Dust2Dust. For BayeSN, we instead calculate posterior distributions on $\tau_E$ by dividing the value of $\tau_A$ by the value of $\mu_{R_V}$ for each step along the chain of posterior samples. Despite this model difference, BayeSN accurately recovers the simulated $\tau_E$ values. Posterior coverage of the truth is also generally good, with $N_{95}$ close to 95 per cent in most cases.

While $\sigma_R$ is recovered perfectly for our Double $R_V$, single $E(B-V)$ simulation, it is noticeable that BayeSN estimates of $\sigma_R$ are generally further than the true simulated values than for either $\mu_{R_V}$ or $\tau_E$. This difference is not large and could not explain historic differences between results of SALT-based analyses and BayeSN. For simulations with $\mu_{R_V}=3.0$, $\Bar{\sigma}_R$ is typically overestimated, the most notable example being 0.77(0.04) for simulated $\sigma_R=0.5$. For simulations with $\mu_{R_V}=2.0$, however, $\Bar{\sigma}_R$ is typically slightly underestimated, the most notable example being 0.41(0.02) for simulated $\sigma_R=0.5$. There are a few reasons why this might be the case. The dust contribution to the $c_{\rm int}$ term in the Dust2Dust simulations (as discussed in Section \ref{sec:LCFitters:subsec:modeldiff}) may be impacting the inferred value of $\sigma_R$. Alternatively, this might result from the influence of the prior on $\sigma_R$ posteriors. $\sigma_R$ is typically less constrained by data than either $\mu_{R_V}$ or $\tau_A$ and a number of previous BayeSN analyses only placed upper limits on $\sigma_R$ \citep[e.g.][]{Thorp21, Thorp24}. In BayeSN, we assume a wide prior $\sigma_R \sim \text{Half-}N(0, 2^2)$. This prior is deliberately wide to cover the range of $\sigma_R$ values that have been proposed in the literature, and in this work we use BayeSN with its typical priors from previous literature rather than tailoring it to our specific application. However, for a simulated value of $\sigma_R=0.5$, with a lack of constraint from the data and such a wide prior an overestimated value of $\sigma_R$ is unsurprising. In these cases where $\sigma_R$ is overestimated, $N_{95}$ is typically 72 per cent showing that these posteriors do not cover the true simulated value as often as we would expect. With { stronger constraint on $\sigma_R$ from data, or perhaps an alternative prior that disfavours large values of }$\sigma_R$, we would expect our posteriors to better cover the truth. Indeed, in cases where BayeSN-estimated $\bar{\sigma}_R$ values are closer to the true value of 0.5 the values of $N_{95}$ are much closer to 95 per cent.

Values of $\Bar{\sigma}_R$ below the true simulated $\sigma_R$ are also not surprising. A number of $\sigma_R$ posteriors are upper limits or have significant density close to 0, which will affect the values of $\hat{\sigma}_{R,i}$ for each simulacrum of the simulation. In these cases, we would expect the value of $\Bar{\sigma}_R$ to be less than the true $\sigma_{R_V}$. Generally, when this occurs $N_{95}$ is close to 95 per cent which demonstrates that these posteriors still have good coverage of the truth. Overall, the small differences we sometimes see between the true $\sigma_{R_V}$ and $\Bar{\sigma}_R$ are not surprising. It is also worth noting that previous dust inference based on SALT has yielded consistently larger values of $\sigma_R$ than BayeSN analyses. The differences we observe in this work between SALT-based sims and BayeSN-estimated $\sigma_R$ values are comparatively small and do not explain these findings. Considering the accurate estimates of $\mu_{R_V}$ and $\tau_E$, it is clear that BayeSN is able to accurately recover dust properties from simulations based on an alternate model.

\subsubsection{{Disentangling Dust Differences from Intrinsic Variation}}
\label{sec:Results:subsec:dust:subsubsec:intdiff}

{One of the main applications of BayeSN within \citetalias{Grayling24}, as well as in Section \ref{sec:Results:subsec:dust} of this work, is to probe the cause of the mass step by disentangling intrinsic differences from differences in dust properties. To test this, we consider a simulation including both differences in dust properties and an intrinsic mass step. As discussed in \ref{sec:Methodology:subsec:simreco}, in this case we opt to use the `intrinsic mag difference' model rather than the `intrinsic SED difference' model. This is because the simulation does contain a constant achromatic mass step, and using this model directly infers a mass step parameter which we can compare to the truth. Results of this analysis are presented in Table \ref{tab:INTReco}.}

{These results demonstrate that BayeSN can indeed disentangle an intrinsic mass step from differences in dust, even when applied to simulations based on a different model. For a true simulated mass step of 0.05 mag BayeSN infers a step of 0.043 mag on average across all 50 simulacra. While this is slightly offset from the truth, small offsets are not surprising given the difference between the model used for simulation and for inference. It appears that this offset in the mass step is instead captured by a slight offset in the inferred $\tau_E$ for SNe Ia in low-mass galaxies. Additionally, the true simulated intrinsic mass step is well covered by the posteriors obtained from BayeSN with $N_{95}=98$ per cent across our 50 simulacra. In fact, the small bias that we observe is comparable to that obtained in a fully SALT-based analysis where the same model was used for both simulation and inference \citep{Popovic21a}.}

{Overall, BayeSN perfectly recovers the true simulated $\mu_{R_V}$ values while finding a significantly non-zero intrinsic mass step between the two populations. This further validates findings presented in \citetalias{Grayling24} and Section \ref{sec:Results:subsec:dust} of this work that there is an intrinsic component of the mass step alongside a potential contribution from dust.}

\begin{table}
    \centering
    \begin{tabular}{c|ccc}
        Parameter & Input Value & Estimated Value (Std. Err.) & $N_{95}$  \\
        \hline
         $\mu_{R_V}$ (HM) & 3.0 & \ORVOEBVHMRVM & \ORVOEBVHMRVMN \\
         $\sigma_{R_V}$ (HM) & 0.5 & \ORVOEBVHMRVS & \ORVOEBVHMRVSN  \\
         $\tau_E$ (HM) & 0.08 & \ORVOEBVHMEBV & \ORVOEBVHMEBVN \\
         $\mu_{R_V}$ (LM) & 3.0 & \ORVOEBVLMRVM & \ORVOEBVLMRVMN \\
         $\sigma_{R_V}$ (LM) & 0.5 & \ORVOEBVHMRVS & \ORVOEBVHMRVSN  \\
         $\tau_E$ (LM) & 0.08 & \ORVOEBVLMEBV & \ORVOEBVLMEBVN  \\
    \end{tabular}
    \caption{Single $R_V$, $E(B-V)$ simulation.}
    \label{tab:1RV1EBVReco}
\end{table}

\begin{table}
    \centering
    \begin{tabular}{c|ccc}
        Parameter & Input Value & Estimated Value (Std. Err.) & $N_{95}$ \\
        \hline
         $\mu_{R_V}$ (HM) & 3.0 & \TRVOEBVHMRVM & \TRVOEBVHMRVMN  \\ 
         $\sigma_{R_V}$ (HM) & 0.5 & \TRVOEBVHMRVS & \TRVOEBVHMRVSN \\
         $\tau_E$ (HM) & 0.08 & \TRVOEBVHMEBV & \TRVOEBVHMEBVN  \\
         $\mu_{R_V}$ (LM) & 2.0 & \TRVOEBVLMRVM & \TRVOEBVLMRVMN  \\
         $\sigma_{R_V}$ (LM) & 0.5 & \TRVOEBVLMRVS & \TRVOEBVLMRVSN  \\
         $\tau_E$ (LM) & 0.08 & \TRVOEBVLMEBV & \TRVOEBVLMEBVN \\
    \end{tabular}
    \caption{Double $R_V$, single $E(B-V)$ simulation.}
    \label{tab:2RV1EBVReco}
\end{table}

\begin{table}
    \centering
    \begin{tabular}{c|ccc}
        Parameter & Input Value & Estimated Value (Std. Err.) & $N_{95}$ \\
        \hline
         $\mu_{R_V}$ (HM) & 3.0 & \TRVTEBVHMRVM & \TRVTEBVHMRVMN \\ 
         $\sigma_{R_V}$ (HM) & 0.5 & \TRVTEBVHMRVS & \TRVTEBVHMRVSN  \\
         $\tau_E$ (HM) & 0.125 & \TRVTEBVHMEBV & \TRVTEBVHMEBVN \\
         $\mu_{R_V}$ (LM) & 2.0 & \TRVTEBVLMRVM & \TRVTEBVLMRVMN \\
         $\sigma_{R_V}$ (LM) & 0.5 & \TRVTEBVLMRVS & \TRVTEBVLMRVSN \\
         $\tau_E$ (LM) & 0.08 & \TRVTEBVLMEBV & \TRVTEBVLMEBVN \\
    \end{tabular}
    \caption{Double $R_V$, $E(B-V)$ simulation.}
    \label{tab:2RV2EBVReco}
\end{table}

\begin{table}
    \centering
    \begin{tabular}{c|ccc}
        Parameter & Input Value & Estimated Value (Std. Err.) & $N_{95}$ \\
        \hline
         $\mu_{R_V}$ (HM) & 3 & \TRVTEBVMBHMRVM & \TRVTEBVMBHMRVMN \\ 
         $\sigma_{R_V}$ (HM) & 0.5 & \TRVTEBVMBHMRVS & \TRVTEBVMBHMRVSN \\
         $\tau_E$ (HM) & 0.125 & \TRVTEBVMBHMEBV & \TRVTEBVMBHMEBVN  \\
         $\mu_{R_V}$ (LM) & 2 & \TRVTEBVMBLMRVM & \TRVTEBVMBLMRVMN  \\
         $\sigma_{R_V}$ (LM) & 0.5 & \TRVTEBVMBLMRVS & \TRVTEBVMBLMRVSN  \\
         $\tau_E$ (LM) & 0.08 & \TRVTEBVMBLMEBV & \TRVTEBVMBLMEBVN \\
    \end{tabular}
    \caption{Double $R_V$, $E(B-V)$, and Gaussian $\beta_{\rm SN}$ distribution simulation.}
    \label{tab:FULLReco}
\end{table}

\begin{table}
    \centering
    \begin{tabular}{c|ccc}
        Parameter & Input Value & Estimated Value (Std. Err.) & $N_{95}$ \\
        \hline
         $\mu_{R_V}$ (HM) & 3 & \INTSTEPHMRVM & \INTSTEPHMRVMN \\ 
         $\sigma_{R_V}$ (HM) & 0.5 & \INTSTEPHMRVS & \INTSTEPHMRVSN \\
         $\tau_E$ (HM) & 0.08 & \INTSTEPHMEBV & \INTSTEPHMEBVN  \\
         $\mu_{R_V}$ (LM) & 2 & \INTSTEPLMRVM & \INTSTEPLMRVMN  \\
         $\sigma_{R_V}$ (LM) & 0.5 & \INTSTEPLMRVS & \INTSTEPLMRVSN  \\
         $\tau_E$ (LM) & 0.08 & \INTSTEPLMEBV & \INTSTEPLMEBVN \\
         $\Delta M_\text{int}$ & 0.05 & \INTSTEPMSTEP & \INTSTEPMSTEPN \\
    \end{tabular}
    \caption{Double $R_V$ simulation with intrinsic mass step.}
    \label{tab:INTReco}
\end{table}

\subsection{Impact of Non-Ia SNe}
\label{sec:simreco:subsec:CC}

Table \ref{tab:CCReco} shows the summary of our BayeSN-estimated dust parameters when we simulate with non-Ia contamination. {These simulations followed the double $R_V$, $E(B-V)$, and Gaussian $\beta_{\rm SN}$ case as presented in Section \ref{sec:simreco:subsec:puresimresults} simulated with additional contaminants}. In the case of placing \textit{no} probability $P_{\rm Ia}$ cuts on our data, we see large biases across all of our parameter estimates, in particular the $\sigma_{Rv}$ components. This is unsurprising given that this sample is dominated by core-collapse supernovae. Indeed, in many cases our BayeSN inference in this case did not converge well given that it was a poor model for this highly contaminated data.
{For our $P_{Ia} > 0$ cut, we find a contamination of 20\%; this decreases to 1\% for the $P_{Ia} > 0.5$ and $P_{Ia} > 0.9$ cuts.}\footnote{{The choice of a cut of 0.5 or 0.9 has little impact on the selected sample as few objects have $P_{Ia}$ values between 0.5 and 0.9.}}

When we impose cuts on $P_{\rm Ia}$\footnote{The ideal solution would be to incorporate $P_{\rm Ia}$ within the likelihood in BayeSN, rather than to cut on $P_{\rm Ia}$. This would require a mixture model treatment of the populations of SNe Ia and contaminants, similar to the Bayesian Estimation Applied to Multiple Species \citep[BEAMS; ][]{Kunz07, Kunz2012} method. This is planned future work for BayeSN.}, the results improve considerably. Even with a small number of contaminants in the data BayeSN is able to recover true simulated dust parameters with reasonable accuracy. As discussed in Section \ref{sec:LCFitters:subsec:modeldiff}, we don't expect completely unbiased parameter recovery between the two models -- even more so with contaminants in the data -- but these results are very promising. Differences between the SALT-simulated and BayeSN-inferred dust parameters are overall similar even when including contaminants, after $P_{\rm Ia}$ cuts are applied.

However, it is very noticeable that the BayeSN estimated value for $\tau_E$ (HM) does not match the true input value, either with $P_{\rm Ia}$ cuts of 0.5 or 0.9. It is important to note that this is not an issue either with BayeSN or the simulation. In fact, this discrepancy arises because simulated SNe Ia with the highest $E(B-V)$ values are removed by the cut on $P_{\rm Ia}$. Even when using a state-of-the-art photometric classifier such as \texttt{SuperNNova}, significant dust reddening on top of a typical SN Ia template will cause misclassification. This is a very important finding and must be taken into account in future analyses of dust properties on photometrically-classified samples. In practice, this sample with $\mu_{R_V}=3$ and $\tau_E=0.125$ has a large impact from dust and is more reddened than previous BayeSN analyses have found with real data \citep[e.g.][]{Thorp21, Mandel22}. The number of simulated SNe Ia removed by the photometric classifier is far reduced for the sample with $\mu_{R_V}=2$ and $\tau_E=0.08$. Nevertheless, our findings demonstrate that great care must be taken to ensure that SNe which are particularly impacted by dust are not misclassified.

The removal of the most highly-reddened SNe means that the sample to which BayeSN was applied had a lower `effective' value of $\tau_E$ than the simulated value of 0.125. Across our 50 simulations the mean simulated $E(B-V)$ value was 0.110 and 0.105 for $P_{\rm Ia}$ cuts of 0.5 and 0.9. After the $P_{\rm Ia}$ cuts are applied the $E(B-V)$ distribution will no longer be exactly exponential, but our BayeSN estimated values of $0.108(0.001)$ and $0.102(0.001)$ are close to these `effective' mean $E(B-V)$ values. Considering only the sample of SNe Ia which BayeSN was applied to, it has performed consistently well at recovering the simulated dust parameters.

Unsurprisingly, the inclusion of contaminants means that our posteriors cover the true simulated dust parameters within the 95 per cent credible region less frequently than for the pure simulated samples discussed in \ref{sec:simreco:subsec:puresimresults}. This is a result of an increased mismatch between the data being simulated and the model being used for inference when including contaminants. Nevertheless, we find that applying BayeSN to a sample with contaminants has a small impact on inferred dust parameters when a cut on $P_{\rm Ia}$ is applied, the impact on $\tau_E$ from the reddest SNe Ia being excluded notwithstanding. Moreover, $P_{\rm Ia}$ is not the only cut we apply to the real DES5YR sample; others are also applied to remove likely contaminants, as described in Section \ref{sec:Data:subsec:sample}. We proceed to apply BayeSN to the real DES5YR sample using a cut of $P_{\rm Ia} > 0.5$.

\begin{table*}
    \centering
    \begin{tabular}{c|ccccccc}
        Parameter & Input Value & \multicolumn{3}{c}{Estimated Value (Std. Err.)} & \multicolumn{3}{c}{$N_{95}$} \\
        & & (No $P_{\rm Ia}$) & ($P_{\rm Ia} > 0.5$) & ($P_{\rm Ia} > 0.9)$ & (No $P_{\rm Ia}$) & ($P_{\rm Ia} > 0.5$) & ($P_{\rm Ia} > 0.9)$ \\
        \hline
         $\mu_{R_V}$ (HM) & 3 & \CCNOCUTHMRVM & \CCSOMECUTHMRVM & \CCMORECUTHMRVM & \CCNOCUTHMRVMN & \CCSOMECUTHMRVMN & \CCMORECUTHMRVMN \\ 
         $\sigma_{R_V}$ (HM) & 0.5 & \CCNOCUTHMRVS & \CCSOMECUTHMRVS & \CCMORECUTHMRVS & \CCNOCUTHMRVSN & \CCSOMECUTHMRVSN & \CCMORECUTHMRVSN \\
         $\tau_E$ (HM) & 0.125$^{*}$ & \CCNOCUTHMEBV & \CCSOMECUTHMEBV$^{*}$ & \CCMORECUTHMEBV$^{*}$ & \CCNOCUTHMEBVN & \CCSOMECUTHMEBVN$^{*}$ & \CCMORECUTHMEBVN$^{*}$\\
         $\mu_{R_V}$ (LM) & 2 & \CCNOCUTLMRVM & \CCSOMECUTLMRVM & \CCMORECUTLMRVM & \CCNOCUTLMRVMN & \CCSOMECUTLMRVMN & \CCMORECUTLMRVMN\\
         $\sigma_{R_V}$ (LM) & 0.5 & \CCNOCUTLMRVS & \CCSOMECUTLMRVS & \CCMORECUTLMRVS & \CCNOCUTLMRVSN & \CCSOMECUTLMRVSN & \CCMORECUTLMRVSN\\
         $\tau_E$ (LM) & 0.08 & \CCNOCUTLMEBV & \CCSOMECUTLMEBV & \CCMORECUTLMEBV & \CCNOCUTLMEBVN & \CCSOMECUTLMEBVN & \CCMORECUTLMEBVN \\
    \end{tabular}
    \caption{Summary of simulated SALT inputs with non-Ia SNe and the BayeSN estimated values.
    \newline
    $^{*}$ Please note, while the simulated input value of $\tau_E$ (HM) was 0.125, the photometric classifier we used misclassified a number of the most highly-reddened SNe Ia and they were removed from the sample by the $P_{\rm Ia}$ cut.
    }
    \label{tab:CCReco}
\end{table*}

\section{Application to DES5YR Sample}\label{sec:Results:subsec:dust}

Having established the ability of BayeSN to generalise to reliable dust inference on simulations based on alternative model, we apply it to our sample of DES5YR light curves. As discussed in Section \ref{sec:Data:subsec:sample}, this work is the first time that BayeSN has been applied to a photometrically-classified sample of SNe Ia. This has a number of implications. On the one hand, as we are not reliant on a reliable spectrum for a classification we should in principle have a more complete sample of SNe Ia. On the other hand, the possibility of a small number of contaminants remains; though we expect this to be mitigated by our cuts on $P_{\rm Ia}$. This cut may, however, cause some of the most highly-reddened SNe Ia to be misclassified and removed from the sample, as discussed in Section \ref{sec:simreco:subsec:CC}.

We test two different BayeSN models on the DES5YR data, for both the Cosmo and BayeDESN samples: one that allows a simple, time-constant peak magnitude difference (intrinsic mag difference) and one that allows for time- and wavelength-dependent changes to the SED as a function of host galaxy mass (intrinsic SED difference). {Both of these models jointly infer differences in dust properties alongside intrinsic differences between each environment.}

\subsection{Intrinsic magnitude difference}
\label{intrinsic_mag_diff}

\begin{figure*}
    \centering
    \includegraphics[width=18cm]{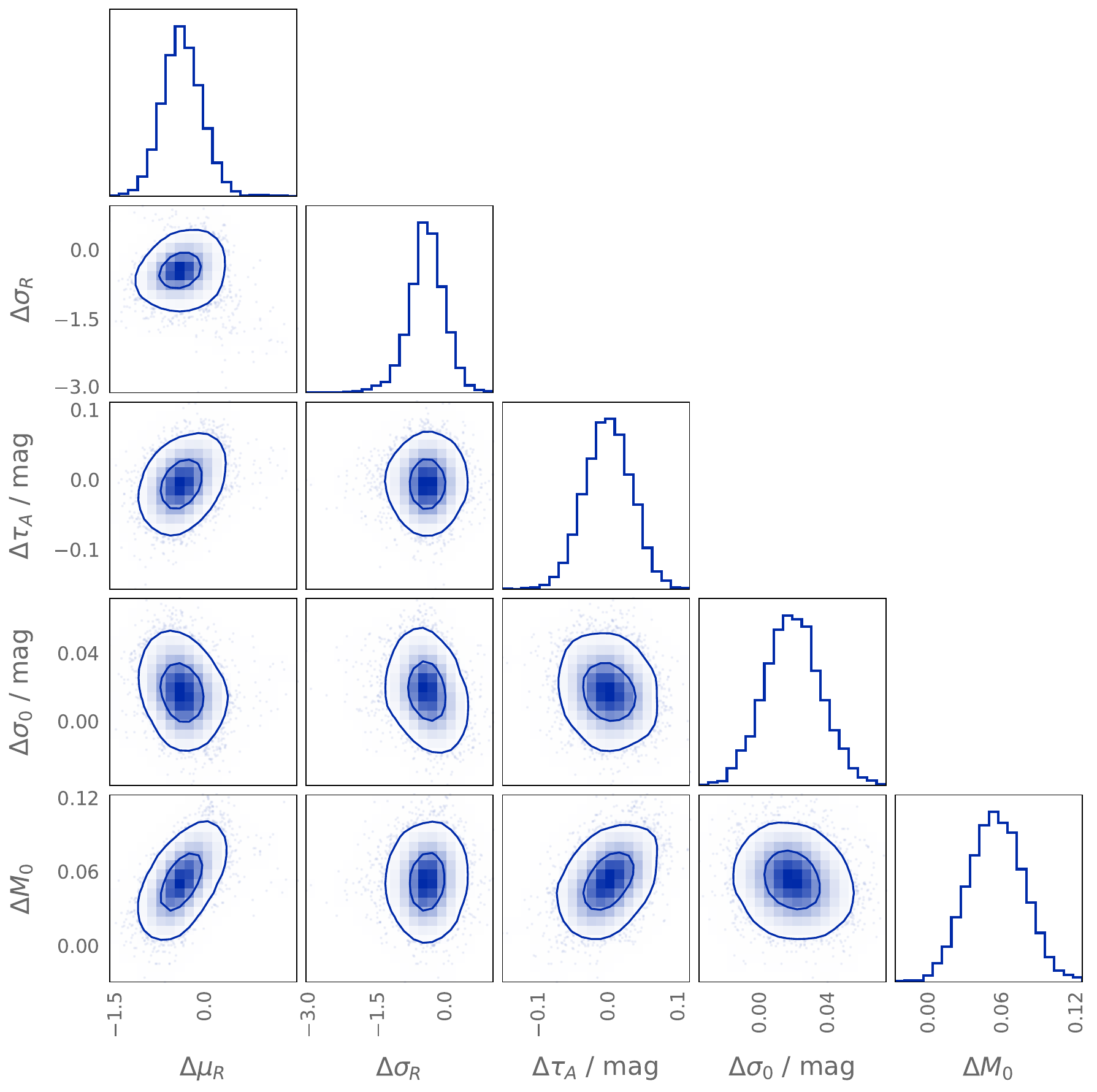}
    \caption{Joint and marginal posterior distributions on the difference between BayeSN hyperparameter values for SNe Ia on either side of the mass step for our Cosmo sample of DES5YR SNe Ia.}
    \label{fig:mag_step_diff_corner}
\end{figure*}

We first consider our `intrinsic mag difference' model as described in Section \ref{sec:LCFitters:subsec:BayeSNmassstep}. In this model, we jointly infer separate host galaxy dust properties with an achromatic, time-constant magnitude difference in the intrinsic SED of SNe Ia on either side of the mass step. Our results for this model are summarised in Table \ref{tab:Mstepresults}. Fig. \ref{fig:mag_step_diff_corner} shows joint and marginal posterior distributions on the difference in parameter values inferred for each mass bin, as well as the intrinsic mass step parameter $\Delta M_0$, for this analysis on our Cosmo sample. For brevity, we do not include a similar plot for the BayeDESN sample as the covariances are very similar.

\begin{table}
    \centering
    \begin{tabular}{ccc}
    Parameter & Cosmo Result & BayeDESN Result \\ 
    \hline
    $\Delta M_0$ & \MstepZMSTM     & \MstepZMSFM \\
    $\mu_{R_V}$ (HM) & \MstepZMSTHMRVM & \MstepZMSFHMRVM \\
    $\mu_{R_V}$ (LM) & \MstepZMSTLMRVM & \MstepZMSFLMRVM 
 \\
    $\Delta\mu_{R_V}$ & \MstepZMSTDELRVM & \MstepZMSFDELRVM \\
    $\sigma_{R_V}$ (HM) & \MstepZMSTHMRVS & \MstepZMSFHMRVS \\
    $\sigma_{R_V}$ (LM) & \MstepZMSTLMRVS & \MstepZMSFLMRVS \\
    $\tau_{A}^*$ (HM) & \MstepZMSTHMEBV  & \MstepZMSFHMEBV \\
    $\tau_{A}^*$ (LM) & \MstepZMSTLMEBV & \MstepZMSTLMEBV 
 \\ 
    $\sigma_{0}$ (HM) & \MstepZMSTHMSIG & \MstepZMSFHMSIG \\
    $\sigma_{0}$ (LM) & \MstepZMSTLMSIG & \MstepZMSFLMSIG \\
    \end{tabular}
    \caption{BayeSN results for the Cosmo and BayeDESN samples when allowing for a simple peak-magnitude offset. \newline
    $^*$ Note the change in convention from $\tau_E$, the input value for SALT simulations, to $\tau_A$, the value directly inferred from BayeSN.}
    \label{tab:Mstepresults}
\end{table}

Our results support the existence of an intrinsic, dust-independent mass step, although at a lower significance than in \citetalias{Grayling24}. Considering only the Cosmo sample (the sub-sample of SNe Ia which were included in the DES5YR cosmology analysis), we infer an intrinsic mass step of $\Delta M_0=0.053\pm0.022$ mag, at a significance of $2.4\sigma$. With the full BayeDESN sample, this is reduced to $0.038\pm0.022$ mag at $1.7\sigma$ significance.

We do see some tentative evidence ($2.4\sigma$) to support a difference in $R_V$ distributions as a function of host galaxy mass. For the Cosmo sample, we infer $\mu_{R_V,HM}=2.57\pm0.20$ and $\mu_{R_V,HM}=2.97\pm0.29$, with the difference $\Delta\mu_{R_V}=-0.41\pm0.33$ at a significance of $1.2\sigma$. This increases, however, when considering the full BayeDESN sample; this yields $\mu_{R_V,HM}=2.61\pm0.14$ and $\mu_{R_V,HM}=3.16\pm0.20$ and $\Delta\mu_{R_V}=-0.55\pm0.23$ at a significance of $2.4\sigma$.

As can be seen in Fig. \ref{fig:mag_step_diff_corner}, the inferred intrinsic mass step $\Delta M_0$ is covariant with the inferred $\Delta\mu_{R_V}$. This result is unsurprising; one of these two effects must explain the mass step present in the sample, hence they trade off against one another.

As discussed in Section \ref{sec:simreco}, inference of $\sigma_{R_V}$ is typically poorly-constrained relative to $\mu_{R_V}$ and $\tau_A$. For $\sigma_{R_V,HM}$, we are only able to obtain 68th (95th) percentile upper limits of 0.36(0.70) and 0.27(0.47) for the Cosmo and BayeDESN samples respectively, which have likely been influenced by the prior. We are able to place more constraint on $\sigma_{R_V,LM}$; $0.76\pm0.35$ and $0.45\pm0.24$ for these two samples.

\subsection{Intrinsic SED difference}

\begin{figure}
    \centering
    \includegraphics[width=8.5cm]{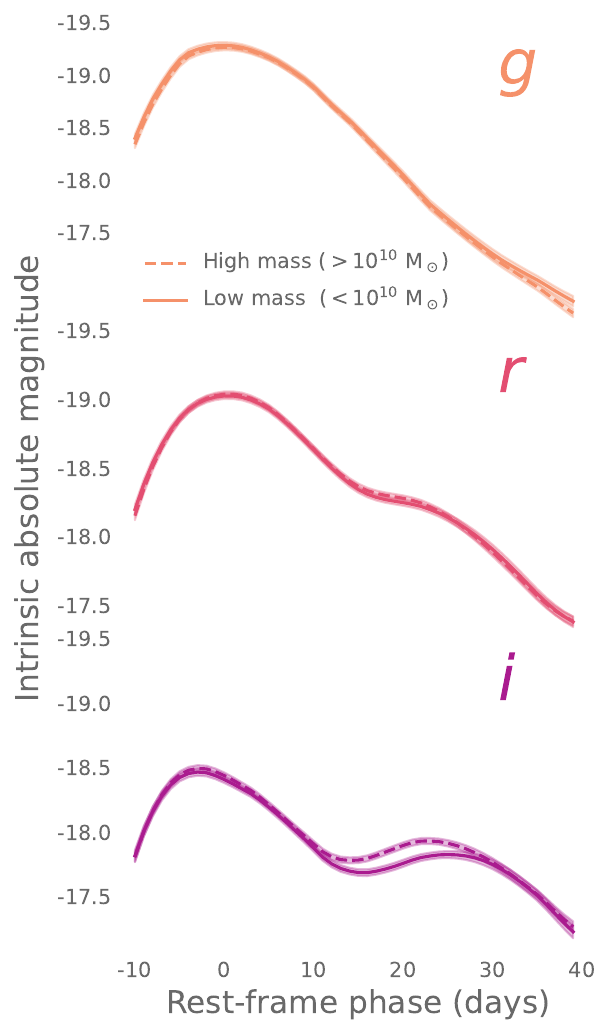}
    \caption{Baseline ($\theta^s=A_V^s=\epsilon^s(t, \lambda)=0$, stretch- and dust-independent) \textit{gri} intrinsic rest-frame light curves for SNe Ia in each mass bin. We calculate these light curves for each step along our posterior chains, and the lines and shaded regions represent the mean and standard deviation across all posterior samples.}
    \label{fig:Msplitlcs}
\end{figure}

We next consider our `intrinsic SED difference' model, which allows for time- and wavelength-dependent differences in the intrinsic SED of SNe Ia on either side of the mass step. Results of this analysis are shown in Table \ref{tab:Msplitresults}, and Fig. \ref{fig:Msplitlcs} shows the dust- and stretch-independent ($\theta^s=A_V^s=\epsilon^s(t, \lambda)=0$) `baseline' intrinsic light curves for SNe Ia in each mass bin inferred when applying this model to our full DES5YR sample. Note that we apply the postprocessing steps outlined in Appendix D of \citetalias{Grayling24} to our MCMC chains when making this comparison. This ensures that we are comparing intrinsic light curves with the same value of $\Delta m_{15}$ in $g$-band.

\begin{table}
    \centering
    \begin{tabular}{ccc}
        Parameter & Cosmo Result & BayeDESN Result \\
        \hline
        $\Delta M_{g,t=0}$ & \ZMSTMgpeak & \ZMSFMgpeak \\ 
        $\Delta M_{i,t=+20}$ & \ZMSTMitwenty & \ZMSFMitwenty \\
        $\Delta(g-r)_\text{peak}$ & \ZMSTMgrpeak & \ZMSFMgrpeak \\
        $\mu_{R_V}$ (HM) & \MsplitZMSTHMRVM & \MsplitZMSFHMRVM \\
        $\mu_{R_V}$ (LM) & \MsplitZMSTLMRVM & \MsplitZMSFLMRVM\\
        $\Delta\mu_{R_V}$ & \MsplitZMSTDELRVM & \MsplitZMSFDELRVM \\
        $\sigma_{R_V}$ (HM) & \MsplitZMSTHMRVS & \MsplitZMSFHMRVS \\
        $\sigma_{R_V}$ (LM) & \MsplitZMSTLMRVS  & \MsplitZMSFLMRVS \\
        $\tau_{A}^*$ (HM) & \MsplitZMSTHMEBV & \MsplitZMSFHMEBV \\
        $\tau_{A}^*$ (LM) & \MsplitZMSTLMEBV & \MsplitZMSTLMEBV \\ 
        $\sigma_{0}$ (HM) & \MsplitZMSTHMSIG & \MsplitZMSFHMSIG \\
        $\sigma_{0}$ (LM) & \MsplitZMSTLMSIG  & \MsplitZMSFLMSIG \\
    \end{tabular}
    \caption{BayeSN results for the Cosmo and BayeDESN samples when allowing for time- and wavelength-dependent SED-level differences between SNe Ia in high- and low-mass host galaxies. The choice of $\Delta M_{i,t=+20}$ is motivated by the secondary peak in the $i-$band as shown in Figure \ref{fig:Msplitlcs}.}
    \label{tab:Msplitresults}
\end{table}

With regard to host galaxy dust, the choice of an intrinsic mag difference or an intrinsic SED difference does not strongly impact our results. For the Cosmo sample, allowing for an intrinsic SED difference yields $\mu_{R_V}$ values of $2.46\pm0.18$ for the high-mass bin and $3.09\pm0.32$ for the low-mass bin. For the full BayeDESN sample, these values are $2.53\pm0.15$ and $3.25\pm0.22$. The differences between these values are $-0.64\pm0.35$ and $0.72\pm0.26$ at significances of $1.8\sigma$ and $2.8\sigma$ respectively. Our inference of $\sigma_R$ and $\tau_A$ is not strongly impacted by the choice of parameterisation of the intrinsic mass step.

With either sample, we do not obtain strong constraints on any intrinsic differences around peak. In $g$-band, our inferred differences in intrinsic absolute magnitude at peak are dominated by their uncertainties; for the Cosmo and full BayeDESN samples respectively, we infer $\Delta M_{g,t=0}$ values of $-0.036\pm0.039$ mag and $-0.019\pm0.045$ mag. These values are both consistent with 0 but also an intrinsic magnitude offset of ${\sim}0.04-0.05$ mag, as inferred using our intrinsic mag difference model and in \citetalias{Grayling24}.

This comparison shows a mass-dependent difference in the intrinsic $i$-band light curve, around the second peak. At 20 days post-peak brightness, this $i-$band difference is $0.136\pm0.045$ mag and $0.139\pm0.039$, at significances of $3.0\sigma$ and $3.6\sigma$, for the Cosmo and BayeDESN sample respectively. This result is in agreement with \citetalias{Grayling24}, which found a similar magnitude offset of $0.099\pm0.022$ mag at $4.5\sigma$ significance. This finding, with an almost entirely independent\footnote{There is some overlap between the DES5YR sample and the DES3YR sample used in \citetalias{Grayling24}} sample of SNe Ia to \citetalias{Grayling24}, provides further evidence to support the existence of an environmental dependence on the intrinsic properties of SNe Ia. We find that, after correcting for the effect of host galaxy dust, two SNe Ia with the same light curve shape around peak but located on either side of the mass step will have significantly different evolution in $i$-band around the second peak. 

There is one notable difference between our findings in this work using the DES5YR sample and \citetalias{Grayling24} which analysed DES3YR, Foundation and PanSTARRS Medium Deep. In this work, when considering $g-r$ colour at peak our results suggest that SNe Ia in high-mass galaxies are intrinsically redder than those in low-mass galaxies. For the Cosmo sample, we find $\Delta(g-r)_\text{peak}=0.042\pm0.016$ mag while for the BayeDESN sample the difference is reduced to $\Delta(g-r)_\text{peak}=0.031\pm0.018$. This is inferred simultaneously with the reduced intrinsic peak luminosity difference discussed above, potentially suggesting that requirements for achromatic luminosity offsets can be relaxed by introducing chromatic color offsets, though large uncertainties in peak luminosity offset demonstrate we cannot draw strong conclusions at present. Following, \citetalias{Grayling24} found that SNe Ia in high-mass galaxies were intrinsically bluer in $g-r$ colour at peak, with $\Delta(g-r)_\text{peak}=-0.022\pm0.010$. As discussed at the beginning of this section, there are fundamental differences between how these samples were observed; DES5YR is a photometric sample, while \citetalias{Grayling24} analysed spectroscopically-confirmed samples. The difference likely arises from greater completeness in the DES5YR sample, although the possible presence of a small number of contaminants in this sample may also partially explain this. Further work is required to make more definitive conclusions about the environmental-dependence of intrinsic colours of SNe Ia.

\section{Discussion and Conclusions}\label{sec:Conclusions}

Previous analyses of the host galaxy dust properties of SNe Ia using SALT and BayeSN have generally found conflicting results. SALT analyses have found large differences in the mean of the $R_V$ distribution, $\mu_{R_V}$, between SNe Ia on either side of the mass step, finding $\Delta\mu_{R_V}\sim1.2-1.6$ \citep{Popovic22, DES5YR}; this approach did not, until recently, allow for achromatic intrinsic differences \citep{Popovic24b}. BayeSN, on the other hand, has supported the existence of an intrinsic contribution to the mass step and predicted $\Delta\mu_{R_V}$ values which are smaller \citepalias{Grayling24} or consistent with zero \citep{Thorp21, TM22}.

In this work we have for the first time {validated the performance of BayeSN by applying it} for inference of dust properties on SALT-based simulations similar to those used in the DES5YR cosmology analysis. Despite the fundamental differences between the models, we find that BayeSN is able to accurately recover simulated dust properties from these SALT-based simulations for a wide variety of permutations of different simulations.

Having established this based on simulations, we then apply BayeSN to real data from the DES5YR sample. We consider two separate parameterisations of an intrinsic mass step with BayeSN and two slightly different samples; one based on the DES5YR cosmology sample after standard cuts and one using the full sample including redder SNe Ia. Neither of these choices substantially impacts our conclusions relating to dust. The largest $\Delta\mu_{R_V}$ we infer is for the full sample when allowing for an intrinsic SED difference on either side of the mass step. This analysis yields $\mu_{R,HM}=2.53\pm0.15$ and $\mu_{R,LM}=3.25\pm0.22$, with $\Delta\mu_{R_V}=-0.72\pm0.26$. Previous analysis of the DES5YR sample using Dust2Dust yielded very different results, with best fit values of $\mu_{R,HM}=1.66$ and $\mu_{R,LM}=3.25$ with $\Delta\mu_{R_V}=-1.59$. 

Overall, we find that BayeSN is able to recover the dust properties from SALT-based simulations, but disagrees with the results from the DES5YR paper \citep{DES5YR}. We propose the explanation that separately fitting the chromatic scatter component \citep[e.g., as in][]{DES5YR} from other observed features of SNe Ia will bias the inferred $R_V$ populations towards a larger $\Delta\mu_{R_V}$, by forcing the different $R_V$ populations to fully explain the observed mass step. A more recent analysis of the DES5YR sample, \cite{Popovic24b}, provides further evidence for this hypothesis: they find that the addition of a grey component to account for scatter and magnitude differences not only improves model fits, but also that a smaller $\Delta\mu_{R_V}$ (more consistent with the results of this paper) is needed when including non-chromatic scatter in the model fit. 

Our BayeSN analysis of the  DES5YR sample supports the existence of environmentally-dependent intrinsic differences across the SN Ia population. When we infer dust properties jointly with an intrinsic mag difference between SNe Ia on either side of the mass step, we find an intrinsic mass step $\Delta M_0=0.053\pm0.022$ for our sample of objects included in the DES5YR cosmology analysis. For our full BayeDESN sample, this is reduced slightly to $\Delta M_0=0.038\pm0.022$ though this value remains consistent with other BayeSN analyses. When we allow for time- and wavelength-dependent intrinsic differences between SNe Ia on either side of the mass step, we find that they have have significantly different intrinsic light curves. After 20 days post-peak, there is an i-band magnitude offset delta magnitude offset $\Delta M_{i,t=20}=0.139\pm0.039$ at a significance of $3.6\sigma$. This result, in tandem with a similar finding in \citetalias{Grayling24} at 4.5$\sigma$ significance on a near independent sample, provides strong evidence that the intrinsic properties of SNe Ia are affected by their environment.

The physical origin of this environmental-dependence of the second peak in $i$-band is uncertain. BayeSN is currently applied only to photometry, leveraging the template from \cite{Hsiao11} for spectral features. As a result, we cannot at present use this method to relate the $i$-band offset to a specific spectral feature, though application of spectra to BayeSN is planned in future. Physical models have previously linked different progenitor properties to the second peak in SN Ia light curves; for example, \cite{Kasen06} found that changes in metallicity impact the second peak while having minimal effect on the first peak. Differences in progenitor properties in different environments, such as metallicity, may explain the host galaxy mass-dependent intrinsic $i$-band difference we find this work. \citet{Deckers24} recently analysed the properties of SNe Ia around the second peak in $r$- and $i$-band, and speculated that metallicity has a stronger impact on the secondary maximum in $i$-band than in $r$-band. This idea would explain why we see such a pronounced offset around second maximum in $i$-band but not $r$-band. Ultimately, further analysis is required to understand what is driving this difference. It is vital for both SN Ia astrophysics and cosmology that future work studies this effect in more detail, linking the $i$-band difference to a physical origin.

One potential cause of time-dependent differences between SNe Ia in different environments is a variable impact of dust during the phase range covered by the model \citep{Forster13}. BayeSN assumes $A_V^s$ and $R_V^s$ are constant for each SN $s$ across the time period covered by the model. Previous studies, however, have suggested that dynamic, or even static, dust can scatter light from the supernova, leading to changes in the observed $E(B-V)$ and $R_V$ over the duration of the SN light curve \citep[e.g.][]{Bulla18a, Bulla18b}, driven by time delays of the SN light entering the observer's line of sight. This could potentially drive time-dependent differences between populations of SNe Ia in different environments. However, these SNe Ia with a time-varying impact of dust are typically found to be particularly red SNe such as SN 2014J (see Table 4 of \citet{Bulla18b}). These would be entirely excluded from cosmology samples such as our Cosmo sample that nevertheless shows this time-dependent intrinsic $i$-band difference.

\subsection{The Impact of Dust on Photometric Classification}

We find that non-Ia contamination biases our inferred dust distributions. Given that non-Ia SNe are more likely to be redder and dimmer, this is not surprising. Without any mitigating measures, we are unable to infer any of our input parameters accurately.
Applying a $P_{\rm Ia} > 0.5$ cut improves our results with the exception of the high-mass $\tau_E$ value. 

The $P_{\rm Ia}$ cut removes a portion of highly-reddened SNe Ia, truncating the tail of the exponential $E(B-V)$ distribution. In practice, the `effective' $\tau_E$ value of the sample after cuts will be less than the true $\tau_E$ value. BayeSN is able to recover this effective value successfully on a sample with a $P_{\rm Ia}$ cut applied; however, this change in the simulated $\tau_E$ population indicates a previously-undiscovered need for improved photometric classification of highly-reddened SNe Ia, and a more streamlined integration of SN~Ia classification into dust-inference pipelines. 

We do not expect this bias in $\tau_E$ to represent a significant bias in the DES5YR results; uncertainty on the dust model is included in the systematic uncertainty budget. However, this bias was not known during the DES5YR analysis, and therefore not included explicitly as a potential systematic.
Nonetheless, this finding underlines the results from \cite{DES5YR, Popovic24a} that astrophysical processes, in particular understanding the impact of dust, remain the largest source of systematic uncertainty in modern SN cosmology.

\subsection{Summary}

In this work we have for the first time applied BayeSN to simulations based on SALT to {examine differences between the two approaches and validate the performance of BayeSN}. We also apply BayeSN to a volume-limited subset of the DES5YR sample of SNe Ia to constrain the environmental-dependence of SN Ia host galaxy dust and intrinsic properties. Our key findings are summarised below:

\begin{itemize}
    \item We verify that BayeSN is able to accurately recover simulated dust properties from SALT-based simulations as used in the DES5YR analysis, despite the fundamental differences between the models. 
    \item We demonstrate that BayeSN is able to successfully disentangle differences in dust properties from intrinsic differences across the SN Ia population when applied to SALT-based simulations.
    \item Following this, we demonstrate that previous differences between inferred values from SALT- and BayeSN-based dust inferences do not stem from diverging treatments of dust.
    \item We find that inferring dust parameters on a simulated DES5YR sample that contains non-Ia contamination is biased  when using the photometric classifier applied for the real DES5YR sample. This bias arises because this classifier has reduced accuracy when applied to the reddest, dimmest SNe Ia rather than the choice of dust-inference method.
    \item Applying BayeSN to our sample of likely SNe Ia from DES5YR, we infer smaller differences in the mean of the host galaxy $R_V$ distribution, $\Delta\mu_{R_V}$, than previous SALT analyses; at most, we find a difference of $0.72\pm0.26$. This difference arises because past SALT analyses did not allow for an intrinsic contribution to the mass step, something more recent analyses have sought to include \citep{Popovic24a}.
    \item With this DES5YR sample, we find evidence to support an intrinsic contribution to the SN Ia mass step. We infer an intrinsic mass step of ${\sim}0.04-0.05$ mag when allowing for a constant magnitude difference between SNe Ia in high- and low- mass galaxies.
    \item We follow up \citetalias{Grayling24} with ${>}3\sigma$ evidence to support an environmental-dependence of the secondary maximum of SNe Ia in the $i$-band when analysing a near-independent sample.
\end{itemize}

\section*{Acknowledgements}

MG is supported by the European Union’s Horizon 2020 research and innovation programme under ERC Grant Agreement No. 101002652 and Marie Skłodowska-Curie Grant Agreement No. 873089. BP acknowledges you, the gentle reader. 

MG and BP would like to thank Kaisey Mandel, Dan Scolnic and Lisa Kelsey for their valuable comments on the manuscript.

%%%%%%%%%%%%%%%%%%%%%%%%%%%%%%%%%%%%%%%%%%%%%%%%%%
\section{Data Availability}

Data used in this article is publicly available with the DES5YR data release from \cite{Sanchez24}, hosted at https://github.com/des-science/DES-SN5YR.



\bibliographystyle{mnras}
\bibliography{research2, matt} % if your bibtex file is called example.bib

%%%%%%%%%%%%%%%%%%%%%%%%%%%%%%%%%%%%%%%%%%%%%%%%%%

%%%%%%%%%%%%%%%%% APPENDICES %%%%%%%%%%%%%%%%%%%%%
\appendix
\section{Manually Cut Supernovae from DES5YR Sample}\label{sec:Appendix}

\begin{figure*}
    \centering
    \includegraphics[width=20cm]{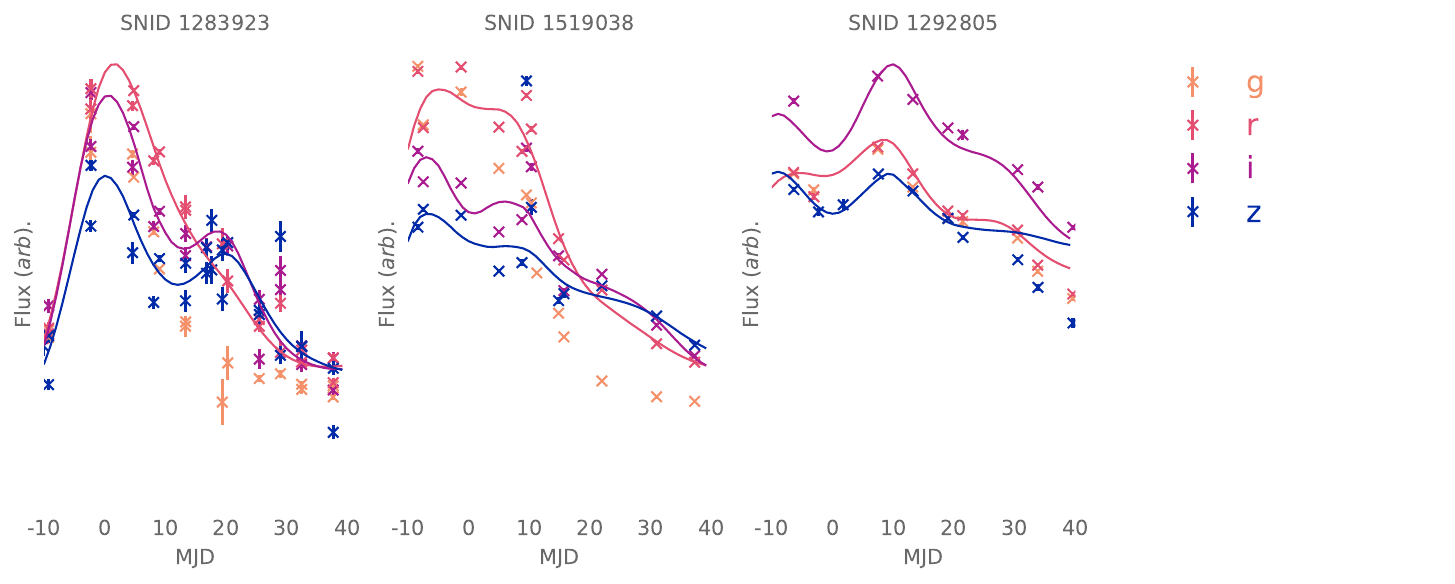}
    \caption{The light-curves that were manually removed from the BayeDESN sample. }
    \label{fig:badlcs}
\end{figure*}
In total, we removed three SNe from our sample which didn't fail any quantitative cuts but caused problems for our inference; the SNIDs of these objects are 1283923, 1519038 and 1292805. The light curves of these objects are shown in Fig. \ref{fig:badlcs}. 

1283923 and 1519038 have multiple epochs of discrepant, high SNR photometry at near-simultaneous phases, as evident in Fig. \ref{fig:badlcs}. This causes general issues when fitting a model due to the strong, contradictory constraints coming from the data. Including these SNe caused our MCMC chains to fail to converge when inferring population-level hyperparameters, therefore we opted to remove this poor quality data.

1292805, meanwhile, was removed because it had an additional peak in its light curve which clearly was not typical for a SN Ia. Fitting this SN with the BayeSN model trained in \citet{Thorp21} led to an extreme inferred $\epsilon^s(t,\lambda_r)$, demonstrating that this SN was not a normal SN Ia. In future, we will establish more quantitative approaches using $\epsilon^s(t,\lambda_r)$ to exclude SNe which are very poorly fit by the model.

%%%%%%%%%%%%%%%%%%%%%%%%%%%%%%%%%%%%%%%%%%%%%%%%%%

% Don't change these lines
\bsp	% typesetting comment
\label{lastpage}
\end{document}